\documentclass{ws-procs10x7}

\newcolumntype{d}[1]{D{.}{.}{#1}}

\makeindex
\begin{document}

\title{BEYOND THE STANDARD MODEL$^\dagger$}

\author{D. I. KAZAKOV}

\address{Bogoliubov Laboratory of Theoretical Physics, Joint Institute
for Nuclear Research, Dubna, Russia\\ Institute for Experimental and Theoretical Physics,
Moscow, Russia,\\ E-mail: kazakovd@theor.jinr.ru}


\twocolumn[\vspace*{-0.8cm}\maketitle\abstract{Review of recent developments in attempts to go beyond the
Standard Model is given. We concentrate on three main unresolved problems: mechanism of
electroweak symmetry breaking, expected new physics at the TeV scale (mainly SUSY) and
the origin of the Dark matter. }]

\section{Introduction}
The Standard Model of fundamental interactions, which is the starting point of all
attempts to look for new physics at high energies, was established as a result of mutual
theoretical and experimental efforts and represents  a solid construction one can be
proud of. Today we face the situation which I would call the HEP paradox: unlike a usual
situation in history when a new theory emerges as a response to unexplained new
phenomena, a modern experiment shows no deviation from the SM and the motivation to go
beyond it comes merely from our desire to explain some features of the SM and our views
on unified theories.

During the last decade there were numerous experimental attempts to find physics beyond
the SM. Search was made for\vspace{-0.3cm}
\begin{itemize}
\item low energy supersymmetry
\item extra gauge bosons
\item axions
\item extra dimensions
\item deviation for the unitarity triangle
\item modification of the Newton law
\item free quarks
\item new forces/particles
\item violation of baryon number
\item violation of lepton number
\item monopoles
\item violation of Lorentz invariance
\item compositeness
\end{itemize}\vspace{-0.3cm}
All of them have failed so far.\\[0.2cm]
$\overline{\phantom{fghffffffffffffffffghjhhkjkk,}}$\\[0.01cm]
$^\dagger$Plenary talk at ICHEP'06, Moscow, July 06

Thus, going beyond the SM one has no hint from experimental data and has to follows one's
own preferences and/or fashion. Still there are some common topics that seem to be of
mutual interest and importance. Below I will concentrate on three main problems of modern
high energy physics.

\section{Problem \#1: Mechanism of Electroweak Symmetry Breaking}
Being very successful in describing three fundamental forces of Nature the SM does not
shed light on the origin of masses. The mechanism of electroweak symmetry breaking is
still not confirmed. So the question is: is it the Higgs mechanism or an alternative one?

\begin{figure}[htb]
\begin{center}
  \includegraphics[height=.30\textheight]{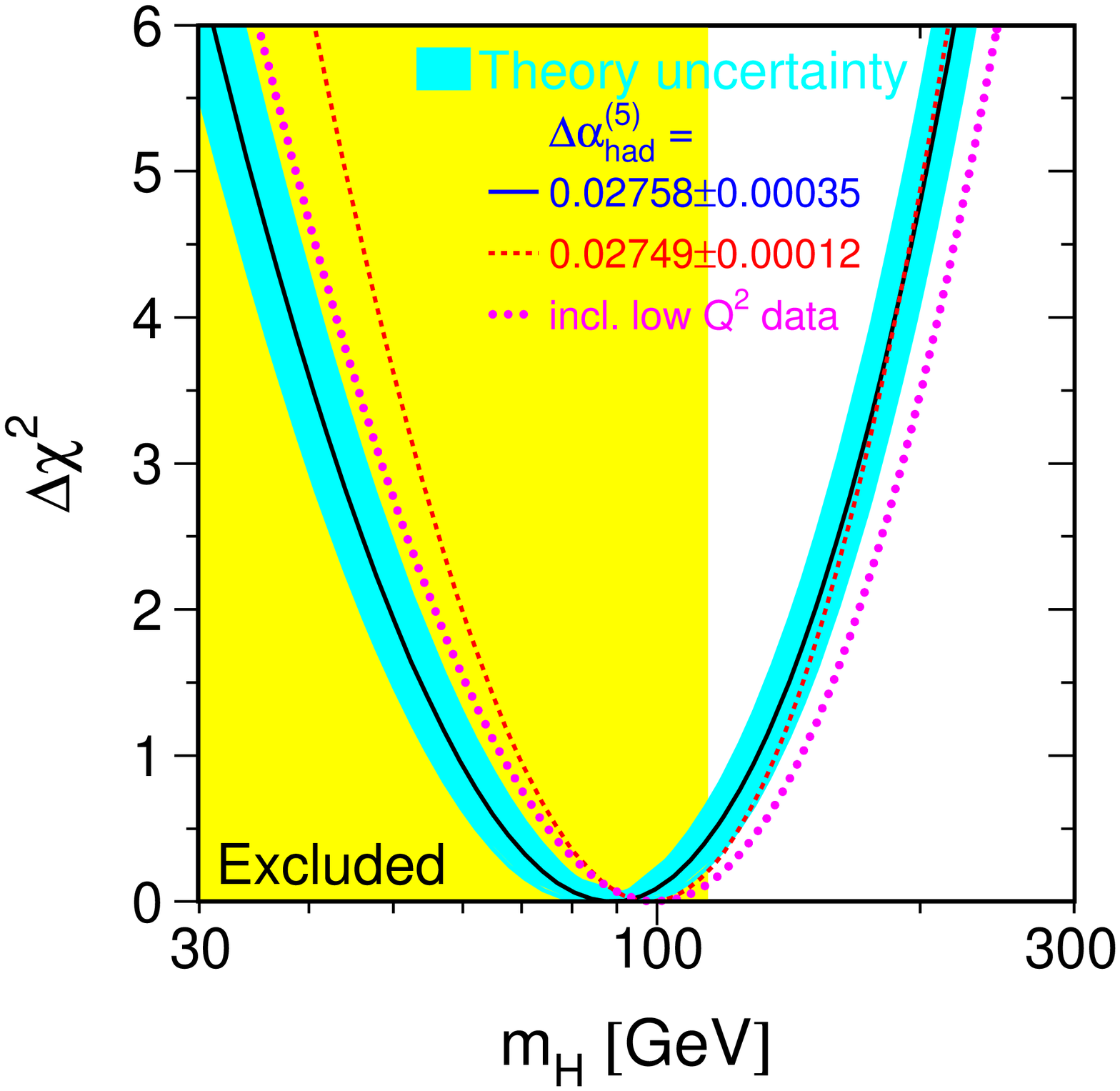}\vspace{-0.8cm}
 \includegraphics[height=.28\textheight]{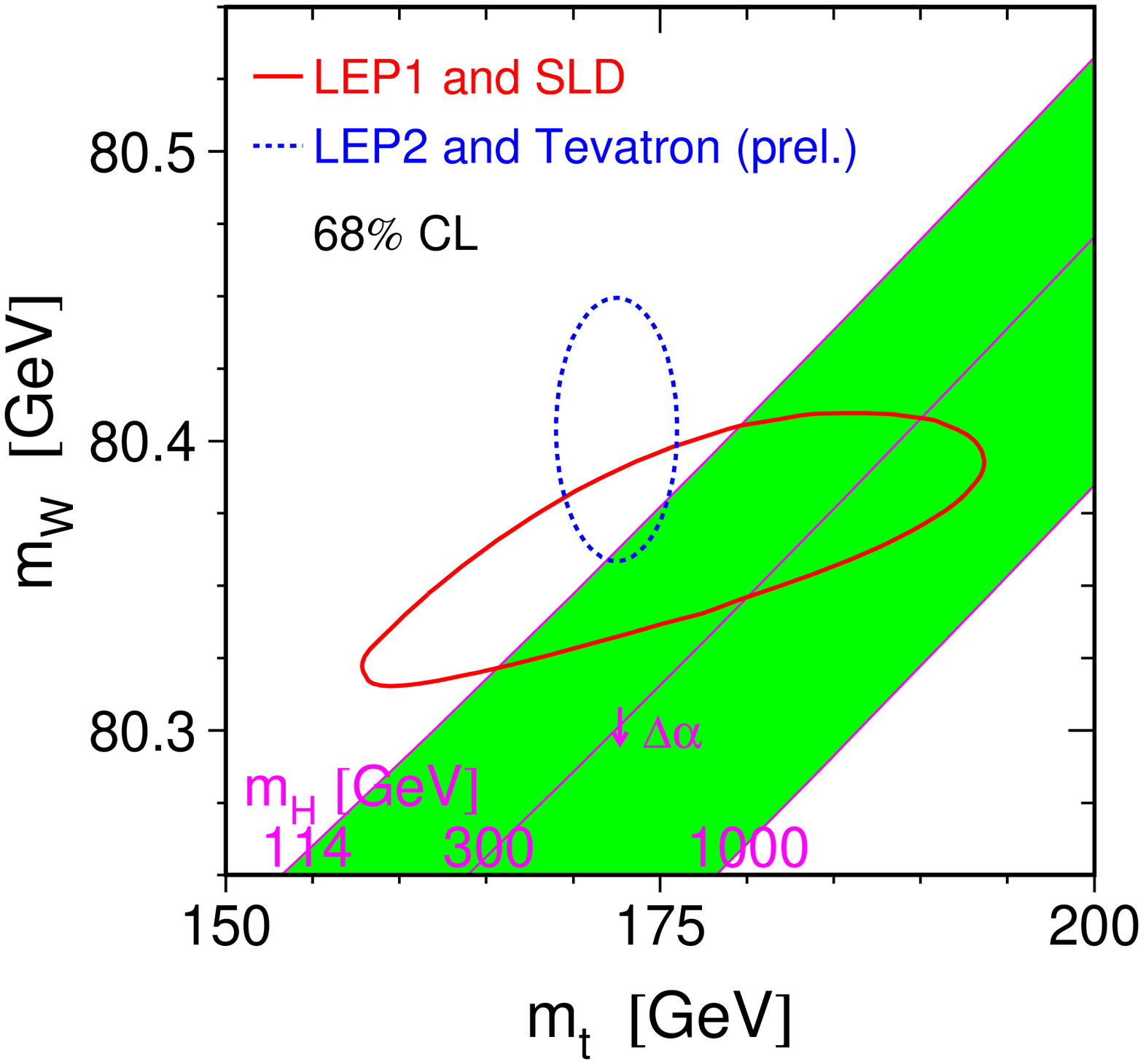}
  \caption{Current indirect limits on the Higgs boson mass}\label{Higgs}
\end{center}
\end{figure}

The standard Higgs boson searches are both direct and indirect. Indirect limits come from
radiative corrections and the direct one comes from the Higgs boson nonobseravation at
LEP II (see Fig.\ref{Higgs})~\cite{limits}
The modern limits on the Higgs boson mass are~\cite{limits}:
 $$\begin{array}{l} M_h=89^{+42}_{-30}\ GeV\ @\ 68\%\ CL \\
 M_h<175\ GeV\ @\ 95\%\ CL \\
\ \ \ \ \ \ \ {\mbox for}\ m_{top}=172.5\ GeV
 \end{array}$$
 So if the Higgs boson is really there, we will see it soon.

 However, one may look for alternatives. They are:

 $\bullet$ Two-Higgs doublet models~\cite{2H}.\\
These models are exploited for many years and have a reach bibliography. The main
difference from the SM is the extra Higgs doublet that introduces new free parameters
including the complex ones which would lead to new sources  of CP violation. Due to the
mixing of states one can make the lightest Higgs boson almost sterile since its
interaction with the Z-boson is suppressed by $\sin(\alpha-\beta)$ that allows one to
have the lightest Higgs mass below 100 GeV without contradicting the modern limits.

$\bullet$ Inert Higgs model~\cite{Barbieri}.\\
 In this model, the inert Higgs doublet has neither vev nor couplings to quarks and
 leptons. After mixing the lightest particle might compose the Dark Matter while the
 usual Higgs boson is heavy ($> 400$ GeV) and does not contradict the precision EW tests.

$\bullet$ Little Higgs models~\cite{ACG}.\\
This class of models represents a new idea of protection of the Higgs mass against
radiative corrections alternative to supersymmetry. The Higgs bosons here are considered
as pseudo-goldstone bosons of some large group similar to the $\pi$-mesons in chiral
theories. In this case, originally the Higgs bosons are massless and obtain their mass
radiatively, but quadratically divergent contributions are not generated and one is left
with the log hierarchy. For this to happen one needs the so-called collective symmetry
breaking and thus a larger gauge group, usually $SU_2\times SU_2$ or $SU_3\times SU_3$.
This leads to new heavy states with masses around 1 TeV. The collider signatures are
similar to SUSY albeit have a different angular dependence due to a different spin
structure~\cite{Perelstein,Birkedal}. To solve the problem of the Dark matter,  one
introduces new parity, called T-parity, similar to R-parity in SUSY models which allows
one to get a stable light particle~\cite{Martin}.

$\bullet$ Twin Higgs model~\cite{Chacko}.\\
It is similar to the Little Higgs model and also treats Higgs boson as a pseudo-goldstone
particle, but has discrete symmetry, twin symmetry, like mirror of L-R symmetry, which
allows one to improve phenomenology. One also has a supersymmetric generalization of this
model~\cite{Chang,Fialkowski}, though the LH model was introduced as an alternative to
SUSY.

$\bullet$ Gauge-Higgs unification models~\cite{Manton}.\\
In this class of models one assumes the existence of extra space-time dimensions. Then
the gauge field has extra components which from the four dimensional point of view can be
considered as scalar particles and one treats the Higgs boson in such a way. One uses
discrete symmetry to protect the Higgs mass from the radiative corrections similar to the
Twin Higgs model. In order to get chiral matter and the Higgs boson in fundamental
representation, one needs an orbifold compactification of extra dimensions. This leads
also to an infinite tower of K-K excitations for W and Z bosons with  masses in the range
of 500 GeV - 1 TeV and extra heavy scalar fields.

$\bullet$ Higgsless models~\cite{Csaki}.\\
In this case one also exploits the idea of extra dimensions with non-flat (warped)
geometry. Electroweak symmetry breaking arise not from the vev of a Higgs field but from
the boundary conditions of a multidimensional field on a four-dimensional brane. The
Lagrangian is symmetric but the boundary conditions are not. This construction allows one
to get W and Z bosons as first K-K excitations together with the infinite tower of states
which are made heavy by warped geometry. Since one has no scalar fields at low energies,
these models are called Higgsless, though scalar fields appear at high energies. What is
essential, unitarity is preserved in this case. There is some problem to get masses for
chiral fermions. To do this, one puts fermions in the bulk and allows a mass term at the
IR vector-like brane which is then translated to chiral fermions on our brane. One has
the usual spin 1 K-K states with the couplings slightly different from the SM, and not to
contradict the EW tests, one needs a heavy compactification scale.

It should be stressed that all these models, contrary to the SM  and its SUSY extensions,
are non-renormalizable and are usually treated as effective low-energy ones.

\section{Problem \#2: New Physics at the TeV scale and search for SUSY}

What is the new physics that is waiting for us at the TeV scale? Is it supersymmetry, or
extra dimensions, or something else? The answer hopefully will come soon. Meanwhile one
should be prepared to discover it. Below I concentrate on SUSY option which is the
mainstream for collider experiments of the last decade and in the near future.

Supersymmetry in this context is understood as various versions of the MSSM which differ
by the way supersymmetry is broken. All of them have different phenomenological
properties and vary in experimental signatures. One usually distinguishes between the
following possibilities:

$\bullet$  MSSM (gravity mediation)~\cite{gravity}.\\
This is the most elaborated version. One usually has 5 universal parameters: $m_0$,
$m_{1/2}$, $A_0$, $\tan\beta$ and $sign(\mu)$ which form the parameter space subjected to
various constraints. Generically squarks and sleptons are relatively heavy (though stop
and sbottom might be light),  gauginos are typically lighter and, depending on the
parameter choice, may decay into leptons besides hadron jets. Production cross-sections
vary with masses but in some regions are big enough for their detection at colliders. The
lightest superparticle (LSP) is usually neutralino which might be light below 100 GeV. In
some cases one may have splitting of masses that leads to metastable particles (gluino,
stau, stop) which may fly through the detector prior to decay. They may even form exotic
states, the so-called R-hadrons where quark or gluino are replaced by their
superpartners, though this possibility usually needs severe fine-tuning.

$\bullet$ MSSM (gauge mediation)~\cite{gauge}.\\
This is the next popular version. Due to the other mechanism of SUSY breaking one has a
different mass spectrum. The LSP is gravitino which might be very light. Since gravitino
interacts only gravitationally, i.e. extremely weakly, the next-to-lightest particle
plays an essential role. It is usually neutralino which decays into photon and gravitino:
$\tilde{\chi}^0_1\to\gamma\tilde G$, i.e. in a final state one gets photons and missing
energy. In this case one may have very long-lived SUSY particles, much longer than in the
gravity mediation case.

$\bullet$ MSSM (anomaly mediation)~\cite{anomaly}.\\
In this case, the mass spectrum of superparticles also differs from that of the universal
SUGRA model.  In particular, the usual universality condition for gaugino masses at the
GUT scale is replaced by the anomaly relation when the gaugino mass ratio is proportional
to the beta function coefficients of the corresponding gauge groups which leads to
inverse hierarchy of gaugino masses. While the RG running these masses can merge at a
lower scale, thus leading to the so-called mirage unification. In this case, like in
gauge mediation scenario, one can also have long-lived charged particles which may decay
inside the detector.

$\bullet$ MSSM (non-universality).\\
Universality assumption introduced in the gravity mediation model reduces the number of
free parameters, thus increasing the predictive power of the model. However, this is not
a physical requirement and can be relaxed. A nonuniversal model naturally allows more
freedom and can satisfy further constraints. A recent review of various possibilities can
be found in H.Baer's talk at SUSY'06~\cite{Baer}.

$\bullet$ NMSSM (singlet extensions).\\
Singlet extensions of the MSSM have their origin in solution of the so-called $\mu$
problem and as a common feature have additional singlet field(s). Due to some additional
freedom here one can relax some constraints and, in particular, increase the value of the
lightest Higgs mass above 120-130 GeV. These models predict also new scalar particles
besides the usual two Higgs doublets. Many models of this type vary in details. Their
summary can be found in V.Barger's talk at SUSY'06~\cite{Barger}.

$\bullet$  MSSM (with R-parity violation)~\cite{parity}.\\
At last, the  R-parity violating models introduce the new lepton or baryon number
violating interactions. If these interactions are suppressed, they do not contradict
modern limits on rare processes but lead to new phenomena. R-parity was invented in order
to stabilize the LSP as a possible candidate for the Dark matter particle, but if LSP is
not stable but long-lived, it can still play its role. One should be accurate though in
applying these new interactions~\cite{Dedes}.

\begin{figure}[htb]
\begin{center}
  \includegraphics[height=.29\textheight]{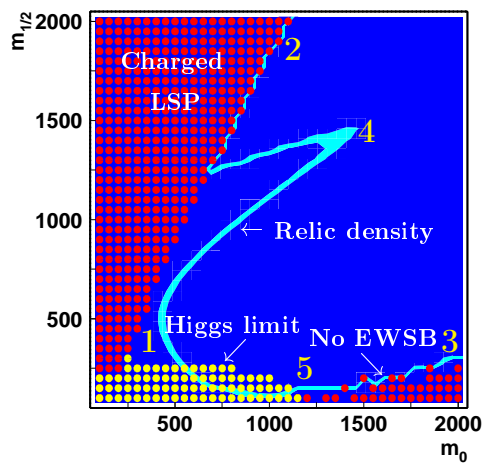}
  \caption{
  The allowed regions of the mSUGRA parameter space: bulk region (1),
  co-annihilation region (2), focus-point region (3), funnel region (4) and EGRET
  region (5). }\label{param}
\end{center}
\end{figure}

Below I consider possible manifestation of supersymmetry at hadron colliders within the
framework of the gravity mediated scenario. The allowed region of the MSSM parameter
space is defined after applying various constraints.  In Fig.\ref{param}~\cite{plane},
the projection of the mSUGRA parameter space onto the $m_0$-$m_{1/2}$ plane is shown for
fixed values of $\tan\beta$ and $A_0$. The left upper corner of the plane is forbidden
due to the requirement of neutrality of the LSP, the left bottom corner is forbidden due
to the Higgs mass limit from LEP and the $b\to s\gamma$ branching ratio, and the right
bottom corner does not allow radiative electroweak symmetry breaking. Accepting the high
experimental accuracy of the measurement of the amount of DM from WMAP, one also gets a
narrow (blue) band allowing the right amount of DM assuming is to be totally made of
supersymmetric particles. Different regions along this band indicated by numbers
correspond to different phenomenological consequences.
\vspace{-0.5cm}

\begin{figure}[htb]
\begin{center}
  \includegraphics[height=.26\textheight]{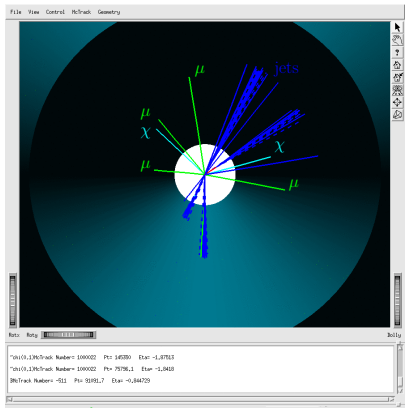}
  \caption{Generation of the process inside the
cylindrical pixel detector in the plane transversely to the beam. One can see 4 muon
tracks (green lines), 2 tracks from neutralino (light blue lines), 4 jets (dark blue
lines) and one long-lived $B$-meson (red line)}\label{detector}
\end{center}
\end{figure}

Looking for superpartners at hadron colliders one should have in mind that they  are
always produced in pairs and then quickly decay creating the ordinary quarks (i.e. hadron
jets) or leptons plus missing energy and momentum. For strong interaction the main
process is the gluon fusion presented in Table 1~\cite{IJ}.
\begin{table*}[t]
\begin{center}
\begin{tabular}{|c|p{0.9cm}|}
\hline Strong Process & final \\
& states \\ \hline
\includegraphics[width=50mm,height=32mm]{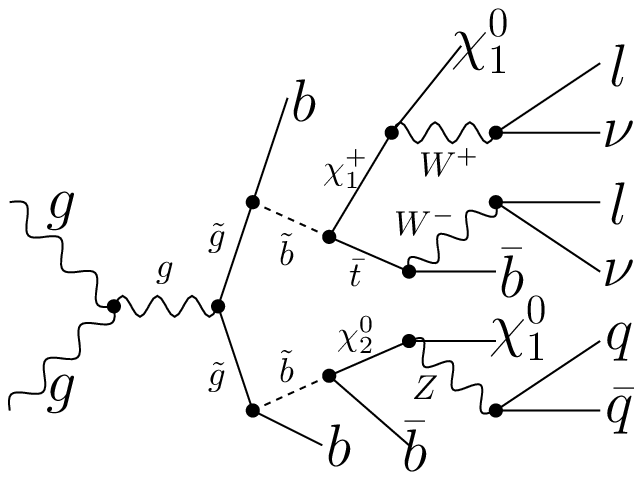}
& \vspace*{-28mm}
\begin{minipage}[t]{1.8cm}
$\begin{array}{c} 2\ell \\ 2\nu \\ 6j \\ \Big/\hspace{-0.3cm E_T}
\end{array}$
\end{minipage} \\ \hline
\includegraphics[width=50mm]{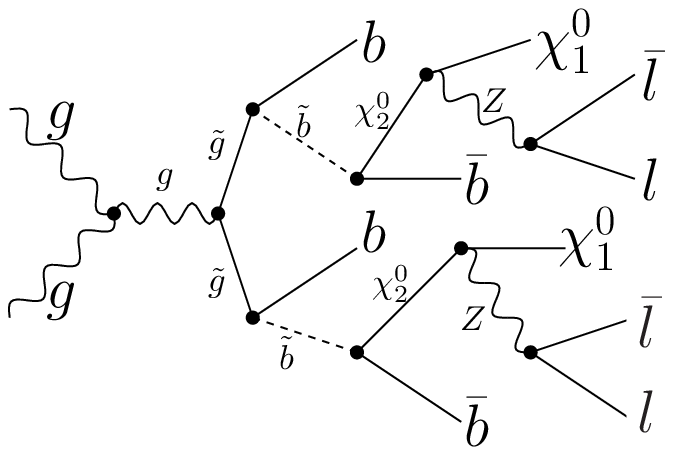}
& \vspace*{-30mm}
\begin{minipage}[t]{1.8cm}
$\begin{array}{c} 4\ell \\ 4j \\ \Big/ \hspace{-0.3cm E_T}
\end{array}$
\end{minipage} \\ \hline
\includegraphics[width=50mm,height=30.5mm]{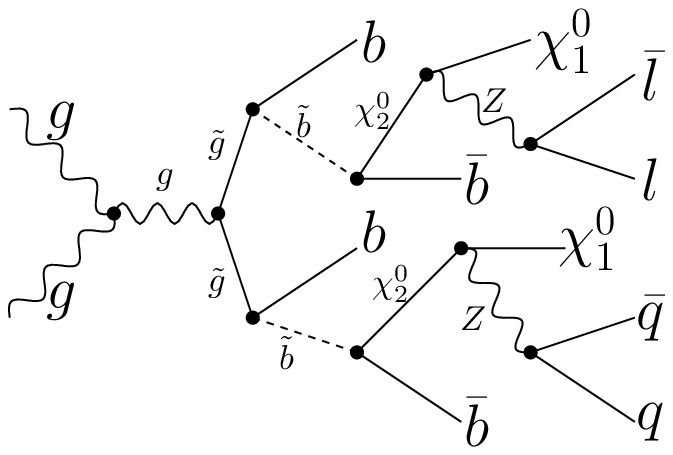}
& \vspace*{-26mm}
\begin{minipage}[t]{1.8cm}
$\begin{array}{c} 2\ell \\ 6j \\ \Big/ \hspace{-0.3cm E_T}
\end{array}$
\end{minipage} \\ \hline
\end{tabular}
\begin{tabular}{|c|p{0.9cm}|} \hline
Weak Process & final \\ & states \\
\hline
\includegraphics[width=40mm,height=32mm]{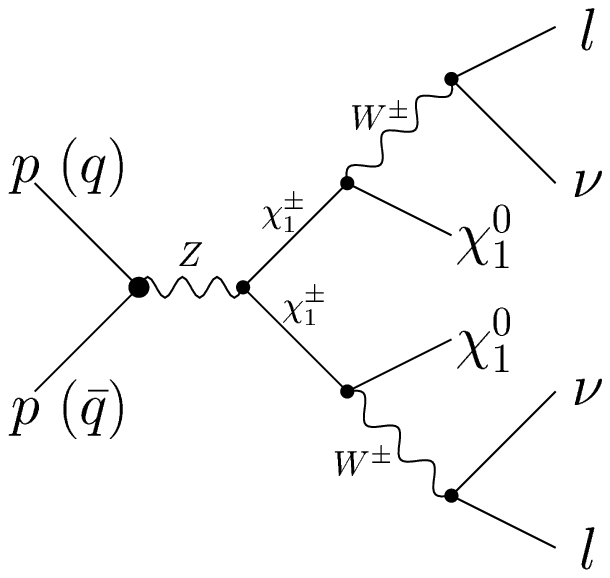}
& \vspace*{-30mm}
\begin{minipage}[t]{1.8cm}
$\begin{array}{c} 2\ell \\ 2\nu \\ \Big/\hspace{-0.3cm E_T}
\end{array}$
\end{minipage} \\ \hline
\includegraphics[width=40mm,height=34mm]{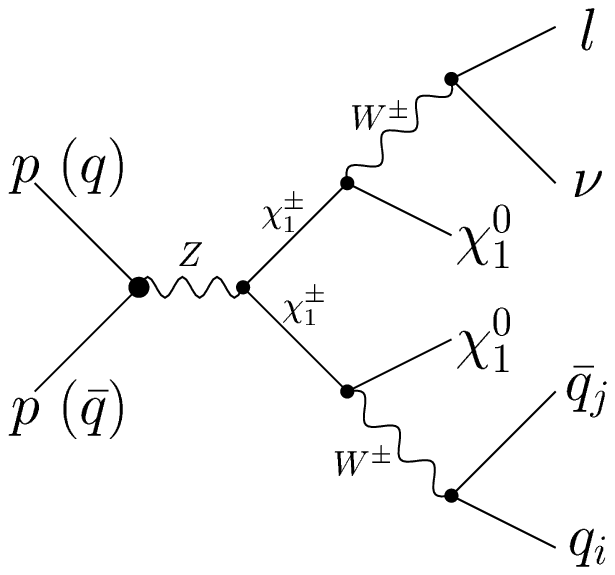}
& \vspace*{-30mm}
\begin{minipage}[t]{1.8cm}
$\begin{array}{c} \ell \\ \nu \\ 2j \\ \Big/\hspace{-0.3cm E_T}
\end{array}$
\end{minipage} \\ \hline
\includegraphics[width=40mm,height=30mm]{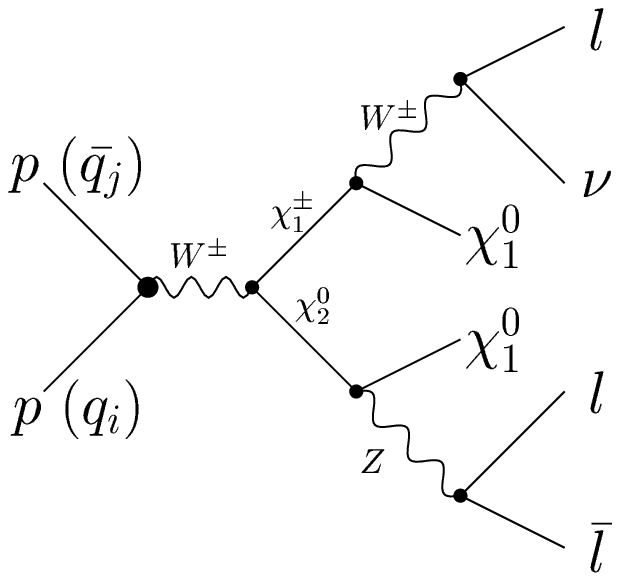}
& \vspace*{-26mm}
\begin{minipage}[t]{1.8cm}
$\begin{array}{c} 3\ell \\ \nu \\ \Big/\hspace{-0.3cm E_T}
\end{array}$
\end{minipage} \\ \hline
\end{tabular}\vspace{0.5cm}

\caption*{\small Table 1: Creation of pairs of gluino (left) and  of the lightest
chargino and the second neutralino (right) with further cascade decay.} \label{tab:n}
\end{center}
\end{table*}
Fig.~\ref{detector} shows a typical event inside the ATLAS pixel detector in the
cylindrical first layer ($R \approx 4$ cm)~\cite{ex}. Particles are produced at the
collision point and decay almost immediately producing hadron jets and muons accompanied
by neutralinos taking away the missing energy and momentum.\vspace{-0.45cm}

Charginos and neutralinos are produced in pairs through the Drell-Yan mechanism  and can
be detected via their lepton decays $\tilde \chi^\pm_1 \tilde \chi^0_2 \to \ell\ell\ell+/
\hspace{-0.3cm E_T}$ (see the Table 1). The main signal of their creation is the isolated
leptons and missing energy.
 The main background in trilepton channel comes from creation of the
standard particles  $WZ/ZZ,t\bar t, Zb\bar b$ è $b\bar b$.

\begin{figure*}[htb]
\begin{center}
  \includegraphics[height=.25\textheight]{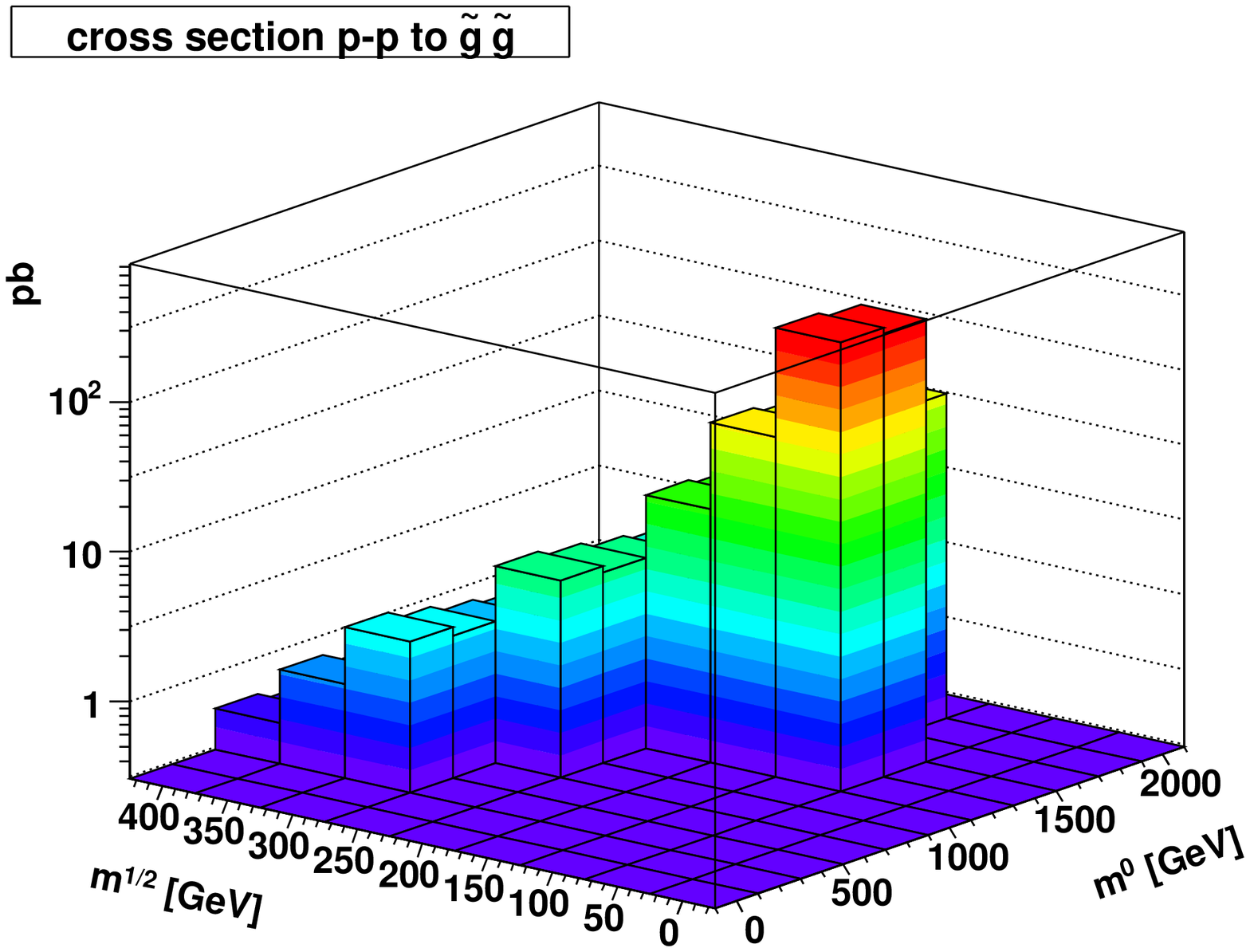}
  \includegraphics[height=.25\textheight]{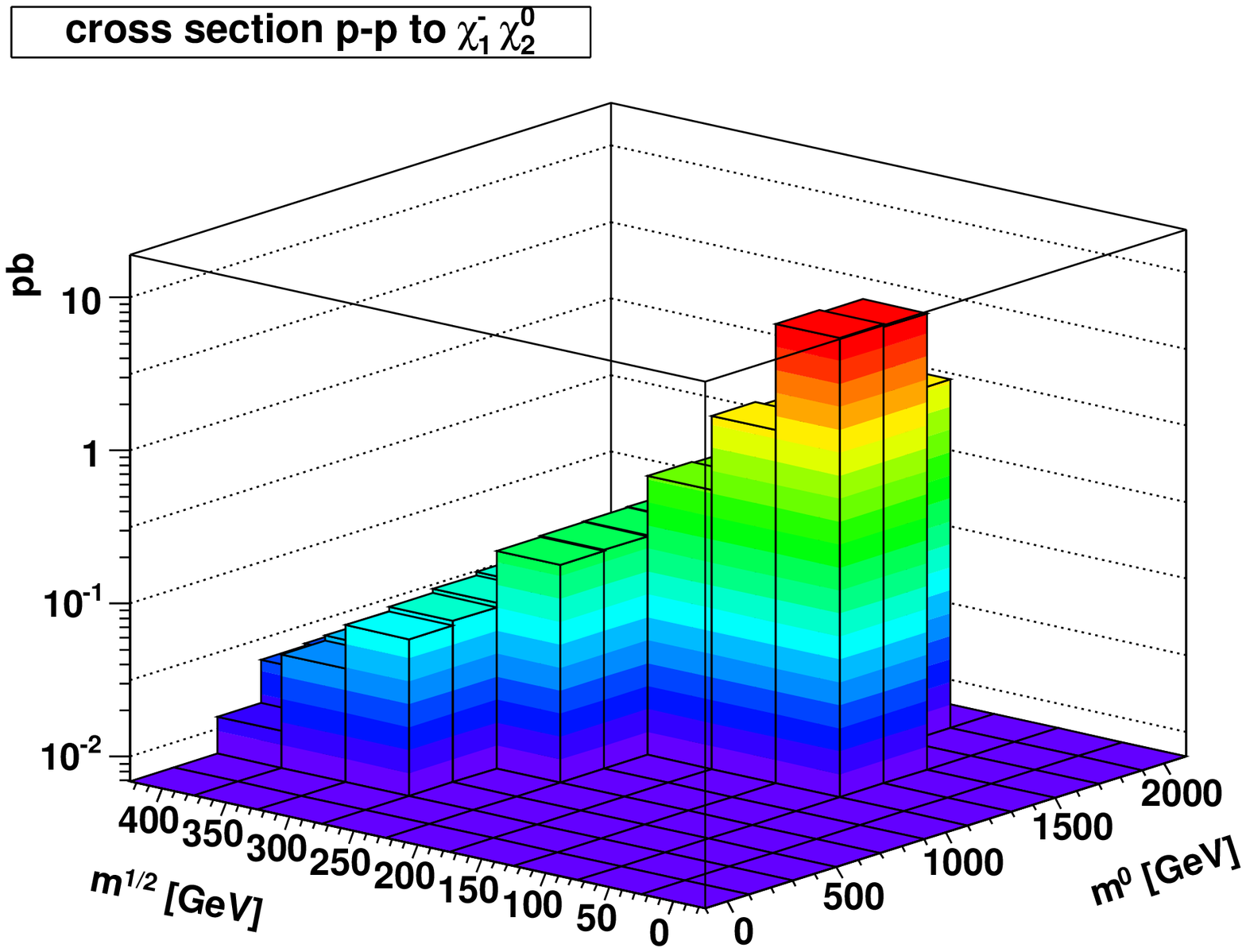}
  \includegraphics[height=.25\textheight]{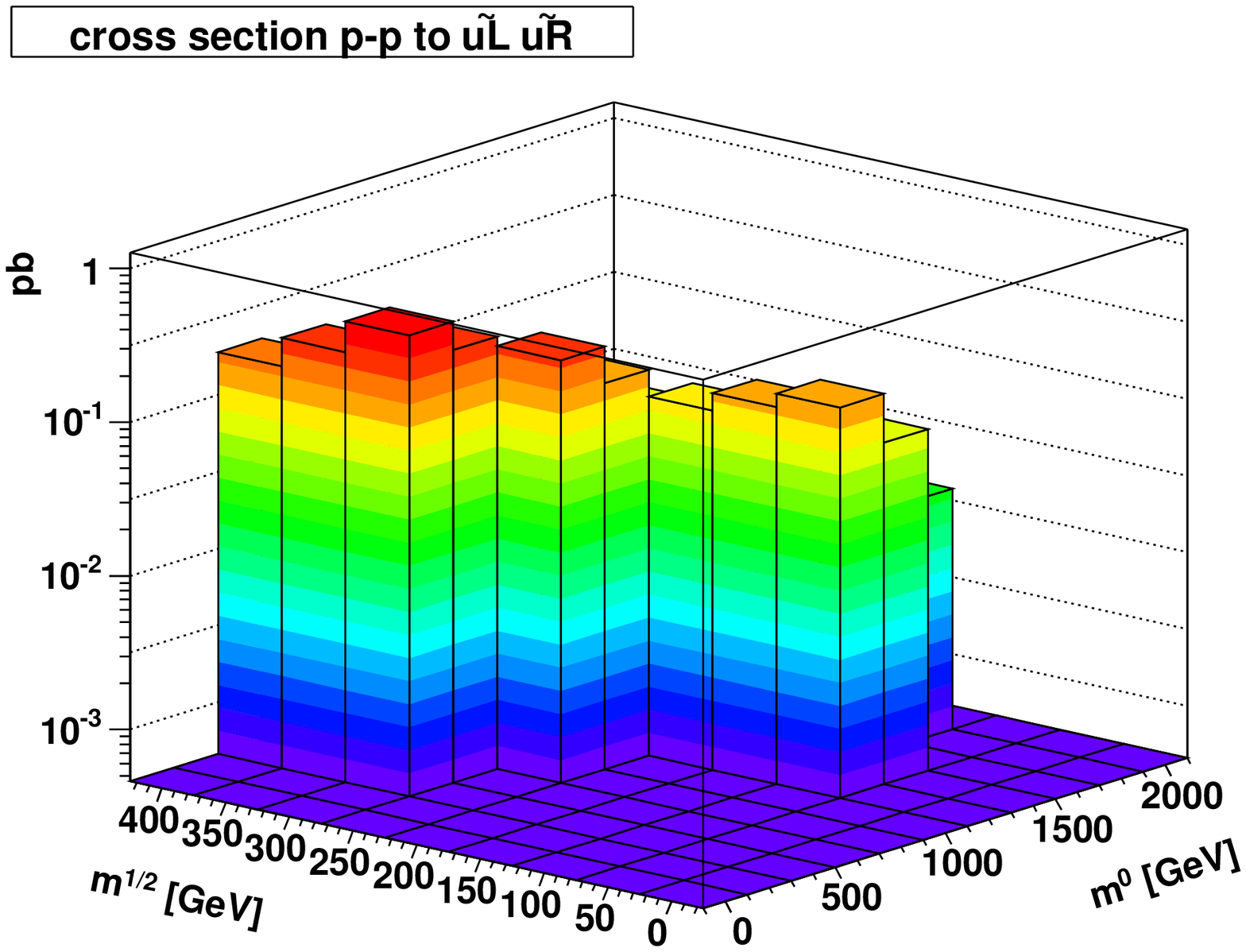}
 \includegraphics[height=.25\textheight]{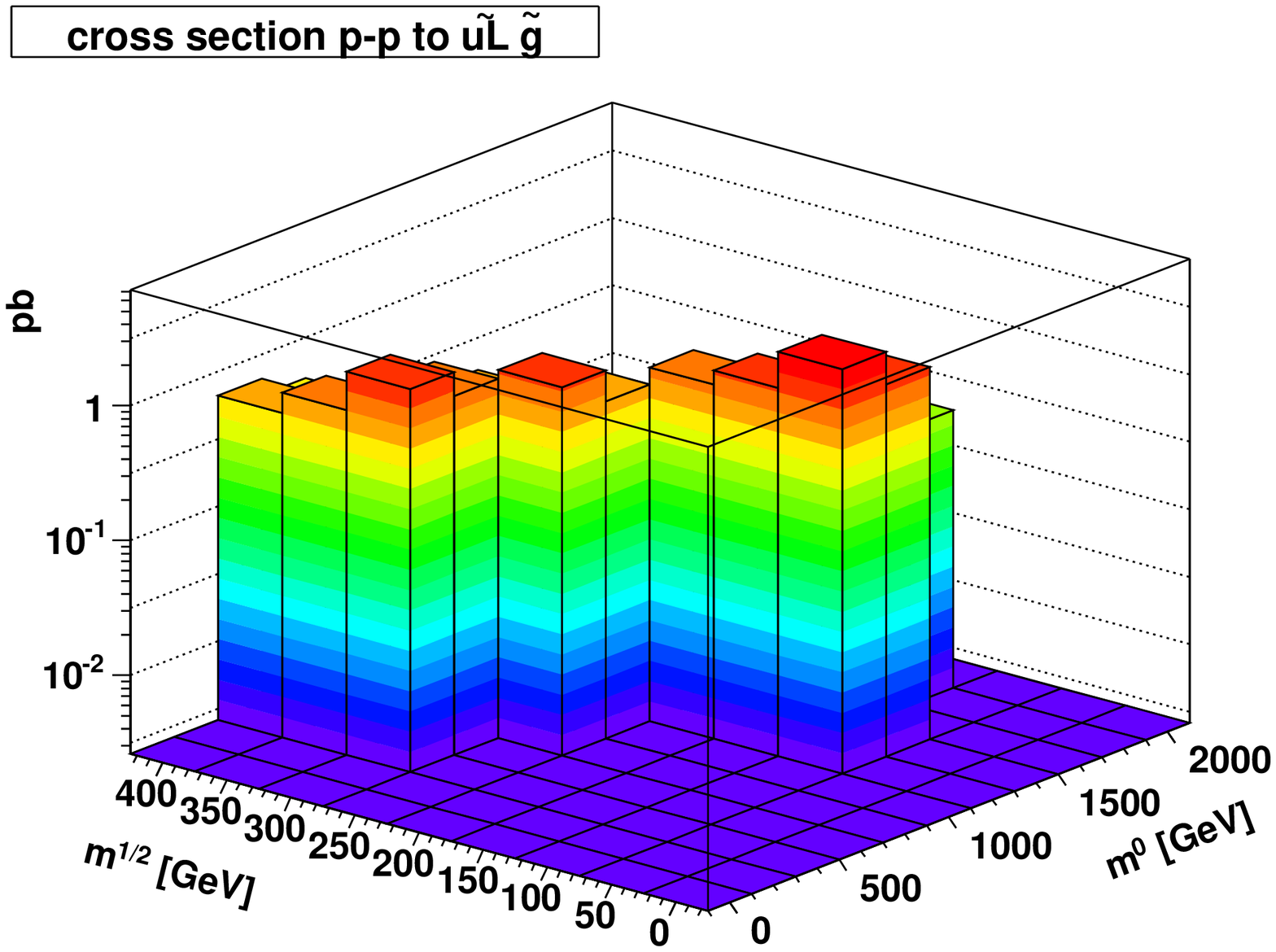}
  \caption{The cross sections of superpartners creation as functions of $m_{1/2}$ and
$m_{0}$ for $\tan\beta=51$, $A_0=0$ and positive sign of $\mu$.}\label{xsec}
\end{center}
\end{figure*}

 The cross-sections at the LHC for various processes in the whole $m_0-m_{1/2}$ plane are
shown in Fig.\ref{xsec}~\cite{ex}. One can see that they vary from a few hundred pb for
gluino production to a few tenth of pb for squark production in the maximum and strongly
depend on the point in parameter space.

\begin{figure}[htb]
\centerline{\includegraphics[width=.45\textwidth]{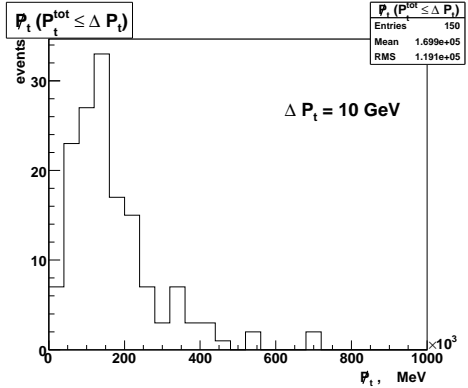}}
 \caption{Total missing transverse
momentum $P_t$ of two neutralinos. Event selection is made assuming that the total $P_t$
of gluino pair is less than 10 GeV. \label{fig:neutralinos}}
\end{figure}

\begin{figure}[htb]
\centerline{\includegraphics[width=.45\textwidth]{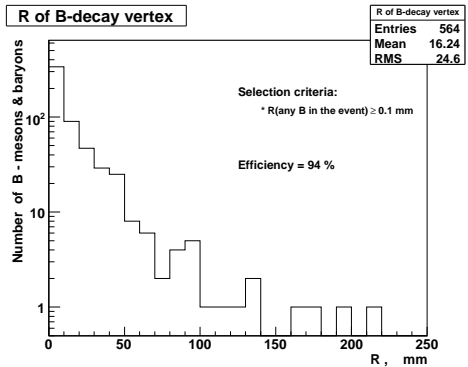}} \caption{The free path of
$B$-hadrons before their decay. \label{fig:freepath}}
\end{figure}

To illustrate the LHC potential in discovering SUSY we consider a gluino production
process with a further cascade decay into jets and muon pairs (process \# 2 from Table
1).
For the choice of parameters corresponding to the region \# 5 in Fig.\ref{param}, the
cross-section of gluino production achieves 13 pb; however, branching ratios into muons
are small and reduce the total cross-section to a few tenth of fb.
In the final state the gluino pair gives 4 $b$-quarks ($b$-jets), 4 muons and a pair of
the lightest stable neutralinos $\tilde\chi^0_1$ giving the high missing transverse
momentum. The jets contain the $B$-hadrons and one may  have four secondary vertices,
which allows one to reduce the background even at the trigger level. Neutralino takes
away quite high transverse momentum.  Fig.~\ref{fig:neutralinos}~\cite{ex} shows the
total transverse momentum of two neutralinos. Careful reconstruction will allow one to
detect such a high loss in the total measured transverse energy.
The $b$-tagging of all $b$-jets appears to be extremely important since the $B$-hadrons
live long enough to move away off the creation point. As a result it allows one to
observe a secondary vertex of the $B$-hadron decay at a certain distance from the primary
beams collision initial vertex and tag hadronic jets from $b$-quarks.
Fig.~\ref{fig:freepath} shows the distribution of the free path of $B$-hadrons provided
all four $B$-hadrons have free paths more than 100 $\rm{\mu m}$ simultaneously~\cite{ex}.
One can see that 94\% of events satisfy this condition.

\begin{figure}[htb]\vspace{-0.4cm}
\centerline{\includegraphics[width=.35\textwidth]{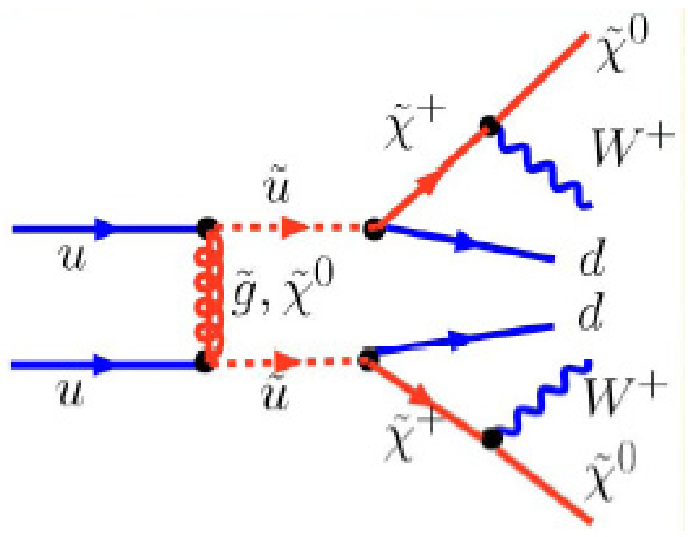}}
\centerline{\includegraphics[width=.35\textwidth]{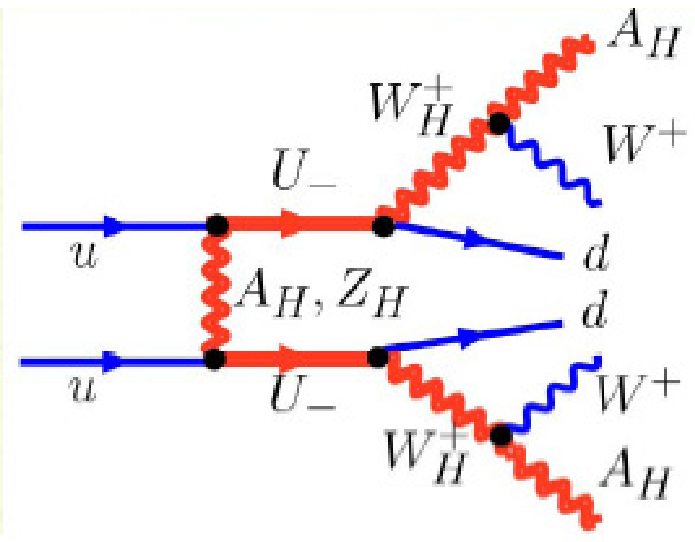}}
 \caption{Comparison of SUSY (up) and
the Little Higgs (down)  missing energy events at colliders \label{LH}}
\end{figure}

It should be mentioned that SUSY event at colliders might be easily mixed up with another
possible new physics. For example, in the Little Higgs models one also has missing energy
events when extra neutral heavy particles, which are present in the spectrum, escape
observation. We show some sample diagrams in Fig.\ref{LH}~\cite{bel}. To distinguish
between these two possibilities one has to carefully study spin correlations and event
rates.

\begin{figure}[htb]
\begin{center}
\includegraphics[width=0.45\textwidth]{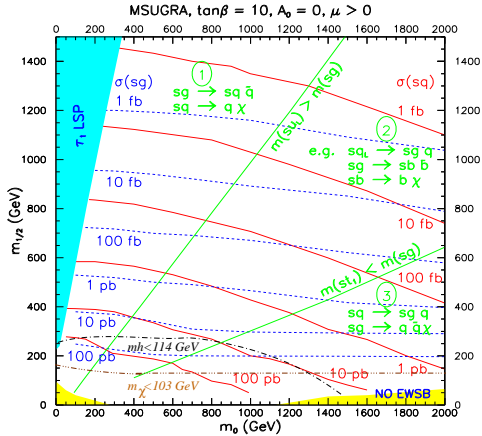}
\end{center}
\caption{Expected range of reach for superpartners in various channels and luminosities
at LHC} \label{fig:LHC1}
\end{figure}

To present the region of reach for the LHC in different channels of sparticle production,
it is useful to consider the same plane of soft SUSY breaking parameters $m_0$ and
$m_{1/2}$. In this case, one usually assumes certain luminosity to be achieved during the
accelerator operation. Fig.~\ref{fig:LHC1}~\cite{LHC} shows these regions of reach in
different channels and different luminosities. The lines of a constant squark mass form
the arch curves, and those for gluino are almost horizontal. The curved lines show the
reach bounds in different channels of creation of secondary particles. The theoretical
curves are obtained within the MSSM for a certain choice of other soft SUSY breaking
parameters.  As one can see, for the fortunate circumstances a wide range of the
parameter space up to the masses of the order of 2~Tev will be examined.

\section{Problem \#3: The Origin of the Dark Matter}

 Cold Dark Matter (CDM) makes up 23\% of the energy of the Universe,
as deduced from the temperature anisotropies in the Cosmic Microwave Background (CMB) in
combination with data on the Hubble expansion and the density fluctuations in the
Universe~\cite{wmap}. In fact, the existence of the Dark matter in the Universe was known
since the late 30's from the motion of clusters of galaxies, rotation curves of stars and
more recently from gravitational microlensing experiments. However, the origin of  DM
remains unclear.

In principle, there are two main options: DM is made of macro objects like brown dwarfs,
dust, micro and macro black holes, etc., or it is made of massive weakly interacting
elementary particles - the so-called WIMPs. The first option is not favorable from
observational data. For the second option we have the following candidates (all of them
beyond the SM):\vspace{-0.2cm}
\begin{itemize}
\item axion (axino) (strong CP)
\item neutralino (SUSY)
\item sneutrino (SUSY)
\item right heavy neutrino
\item gravitino (SUSY)
\item heavy photon (LH)
\item heavy pseudo-goldstone (LH)
\item light sterile Higgs (Inert H)
\end{itemize}\vspace{-0.25cm}
One may probably add to the list. None of them is observed so far.

There are two ways to detect the DM: direct and indirect. Direct DM detection assumes
that the DM particle hits the Earth and interacts with nucleons of a target. With deep
underground experiments one may hope to detect such an interaction. There are several
experiments available: DAMA, Zeplin, CDMS and Edelweiss. Only DAMA claims that they see
the effect in seasonal modulation with fitted mass of around 50 GeV~\cite{DAMA}. All the
other experiments do not see it. The reason might be in different methodic and different
targets, since the cross-section of nucleus-DM interaction depends on a spin of a
nucleus. Still, today we do not have convincing evidence of the DM interaction.

Indirect detection is aimed to look for a secondary effect of DM annihilation in the form
of extra gamma rays and charged particles (positrons and antiprotons) in cosmic rays.
These particles should have an energy spectrum which reflects their origin from
annihilation of massive particles and is different from the background one of the known
sources. Hence, one should have some shoulders in the cosmic ray spectrum. There are
several experiments of this type: EGRET (diffuse gamma rays) to be followed by GLAST,
HEAT and AMS01 (positrons) to be followed by PAMELA, BESS (antiprotons) to be followed by
AMS02. All of these experiments see some deviation from the background in the energy
spectrum, though experimental uncertainties are rather big.

One of the most popular CDM candidates is the neutralino, a stable neutral particle
predicted by supersymmetry~\cite{lspdm}.  In a recent paper~\cite{us} we showed that the
observed excess of diffuse Galactic gamma rays has all the properties of the $\pi^0$
decays of  mono-energetic quarks originating from the annihilation of the DM.

\begin{figure}[h]\vspace{-0.55cm}
  \includegraphics[height=.272\textheight]{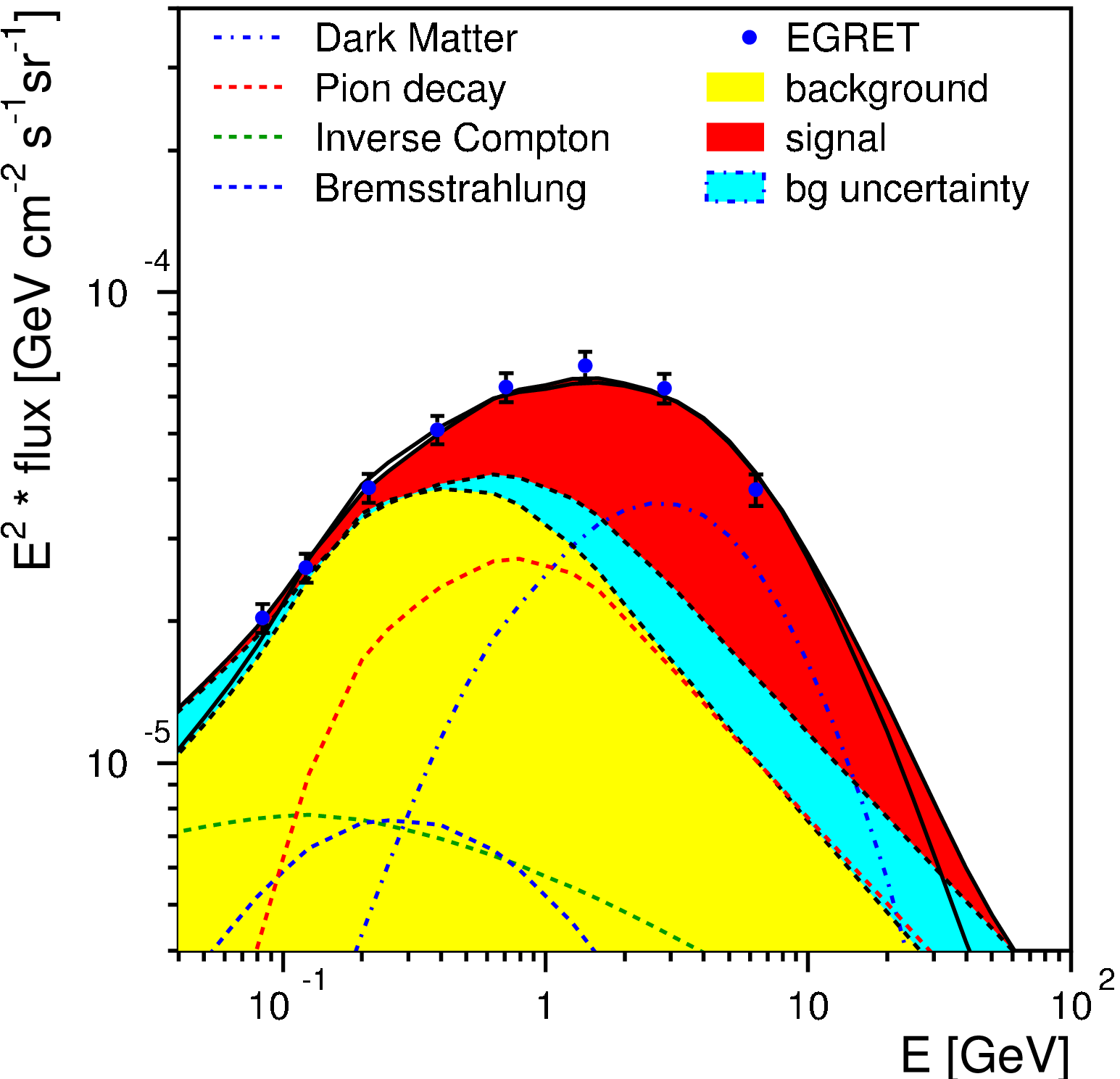}\vspace{-0.4cm}

  \includegraphics[height=.30\textheight]{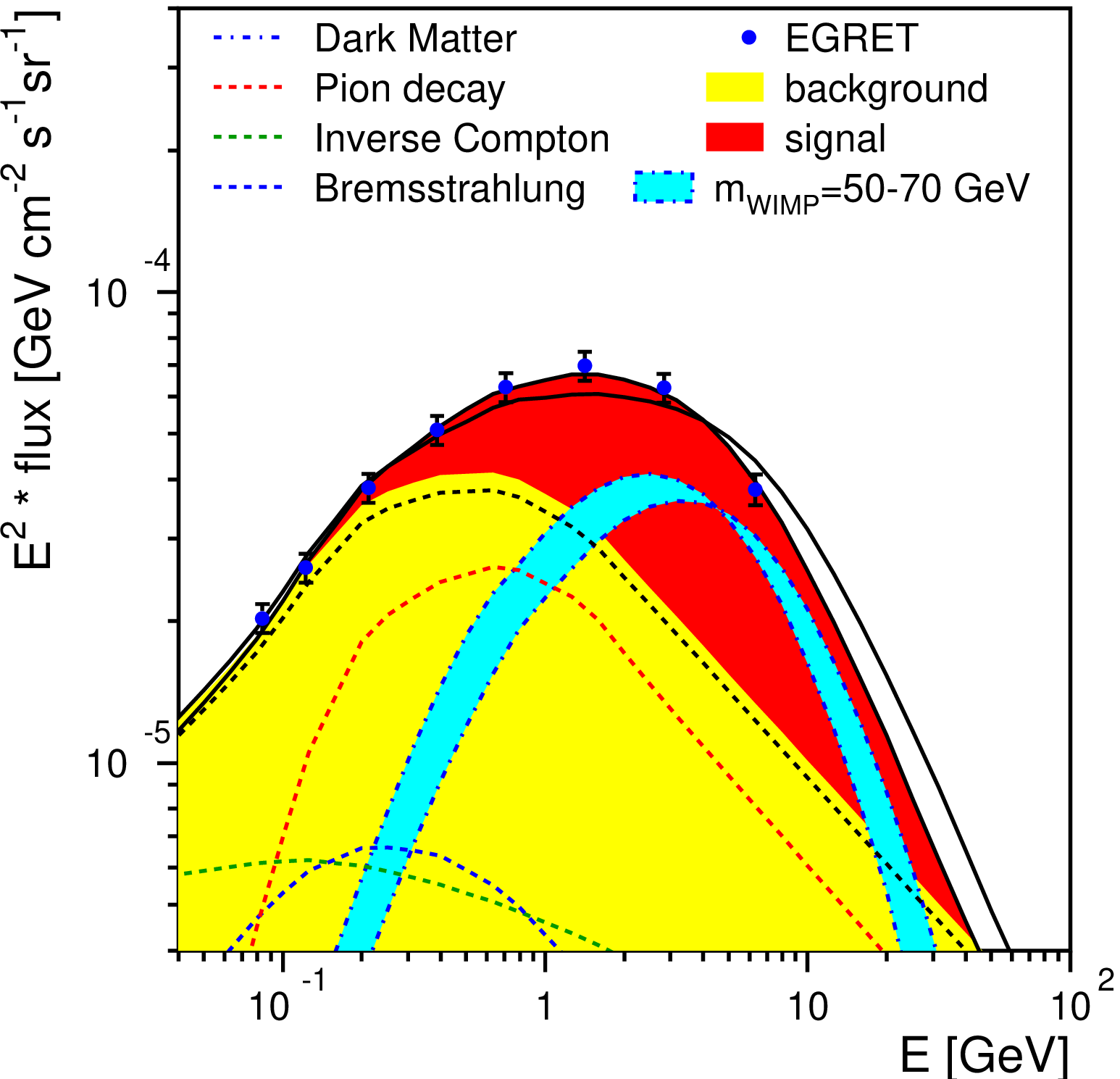}
  \caption{The EGRET gamma ray spectrum fitted with DM annihilation for
  $m_0 =1400$ GeV, $m_{1/2} =175$ GeV, $\tan\beta=51$.  The possible variation of the
  background (blue shaded area above) is  not enough to accommodate the EGRET signal.
 The variation of the WIMP mass between 50 and 70 GeV  shown by blue shaded
  area below is  allowed by the EGRET data with the conventional background \label{DMA}}
\end{figure}

The spectral shape of the diffuse Galactic gamma rays has been  measured by the EGRET
satellite in the range 0.1 - 10 GeV. It allows an independent analysis in many different
sky directions. Comparing the background with the EGRET data shows that above 1 GeV there
is a large excess of gamma rays which reaches more than a factor of two towards the
Galactic centre. However, fitting the background together with the DMA yields a perfect
fit in all sky directions for a DM particle mass around 60 GeV as shown in Fig.\ref{DMA}.
%

The distribution of Galactic diffuse gamma rays measured by EGRET over all sky directions
allows one to reconstruct the profile of DM in our galaxy and to explain the peculiar
shape of rotation curve of stars~\cite{us}.

\begin{figure}[ht]\vspace{-0.3cm}
  \includegraphics[height=.28\textheight]{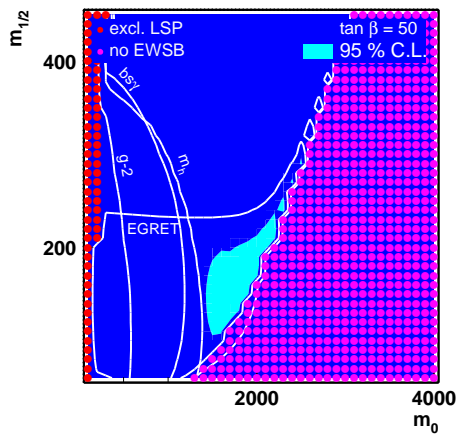}
  \caption{The allowed regions of the mSUGRA parameter space with account of EGRET
  data. The light shaded area (blue) indicates the 95\% C.L. parameter
  range  allowed by EGRET data, the individual constraints have been
  indicated by the lines and dots.}\label{msugra}
\end{figure}

 This intriguing hint of DMA is compatible with supersymmetry,  assuming that the
EGRET excess originates from the annihilation of the stable, neutral lightest
supersymmetric particles, the neutralinos. Their mass is then constrained to be between
50 and 100 GeV ($m_{1/2}$ between 125 and 175 GeV) from the EGRET data, which strongly
constrains the masses of all other SUSY particles, if mass unification at the GUT scale
is assumed. Combining the EGRET data with other constraints, like the electroweak
precision data, Higgs mass limits, chargino mass limits, radiative electroweak symmetry
breaking and relic density leads to a very constrained allowed region of the SUSY
parameter space shown in Fig.\ref{msugra}. Choosing a point in this region gives the SUSY
mass spectrum with light gauginos and heavy squarks and sleptons (see the Table 2).

The lightest neutralino
is an almost pure bino in this case meaning that the DM is a superpartner of the CMB.

Scanning over the allowed region of Fig. \ref{msugra} and demanding an LSP mass above 50
GeV requires $\tan\beta$ to be in the range of 50 to 55~\cite{pl}. The strong dependence
of the relic density on $\tan\beta$ originates from the strong dependence of the
pseudoscalar Higgs mass.

\begin{table}[ht]\vspace{-0.2cm}
 {\begin{tabular}{|c|c|}
   \hline
  {\bf Particle} & {\bf Mass [GeV]} \\
   \hline
   $\tilde \chi^0_{1,2,3,4}$ & 64, 113, 194, 229 \\
   $\tilde \chi^\pm_{1,2},\tilde{g}$ & 110, 230, 516 \\
   $\tilde u_{1,2}=\tilde c_{1,2}$ & 1519, 1523 \\
   $\tilde d_{1,2}=\tilde s_{1,2}$ & 1522, 1524 \\
   $\tilde t_{1,2}$ & 906, 1046 \\
   $\tilde b_{1,2}$ & 1039, 1152 \\
   $\tilde e_{1,2}=\tilde \mu_{1,2}$ & 1497, 1499 \\
   $\tilde \tau_{1,2}$ & 1035, 1288 \\
   $\tilde \nu_e, \tilde \nu_\mu, \tilde \nu_\tau$ & 1495, 1495, 1286 \\
   $h,H,A,H^\pm$ & 115, 372, 372, 383 \\
   \hline
   Observable & Value \\
   \hline
   $Br(b\to X_s\gamma)$ & $3.02 \cdot 10^{-4}$ \\
   $\Delta a_\mu$ & $1.07\cdot 10^{-9}$ \\
   $\Omega h^2$ & 0.117 \\
   \hline
  \end{tabular}}\vspace{0.3cm}

\caption*{\small Table 2: SUSY Particle spectrum at the EGRET point: $m_0=1500$ GeV,
$m_{1/2}=170$ GeV, $A_0=0$, $\tan\beta=52.2$, Sign $\mu=+$ \label{t1}}
\end{table}\vspace{-0.3cm}

\begin{figure}[t]
 \includegraphics[height=.28\textheight]{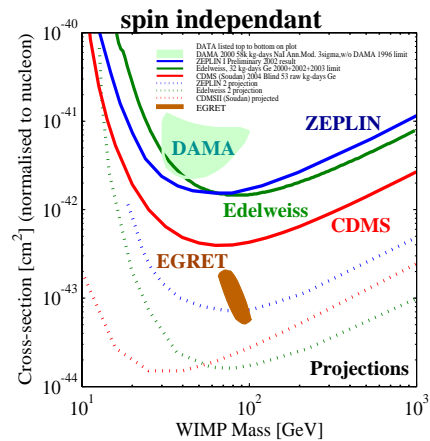}
  \caption{Cross-section of DM nucleus interaction versus the DM particles mass
  and discovery reach of various experiments}\label{DMdirect}
\end{figure}

Given the mass of neutralino one can calculate the cross-section of its interaction with
the nucleus and compare it with the reach of direct search experiments. This comparison
is shown in Fig.\ref{DMdirect}~\cite{dmsearch}. One can see that the cross-section is two
orders of magnitude smaller than the present experimental reach, but will be covered soon
by the forthcoming experiments.\vspace{-0.1cm}

\section{Conclusion}

Future will show us whether we are on the right track and discoveries are waiting for us
round the corner or some unexpected reality is going to emerge. Stakes are high. I would
like to conclude with quotation from St.John "Blessed are those who believe and yet have
not seen"\cite{SJ}.

\end{document}